\documentclass{iau}
\usepackage{graphicx}
\usepackage{xspace}

\newcommand{\apj}{\textit{ApJ}} 
\newcommand{\apjs}{\textit{ApJS}} 
\newcommand{\mnras}{\textit{MNRAS}} 
\newcommand{\nat}{\textit{Nature}} 
\newcommand{\aap}{\textit{A\&A}}
\newcommand{\aj}{\textit{AJ}}
\newcommand{\hi}{H\,{\sc i}\xspace}
\newcommand{\ovi}{O\,{\sc vi}\xspace}
\newcommand{\xiag}{$\xi_{\rm ag}$\xspace}
\newcommand{\xiaa}{$\xi_{\rm aa}$\xspace}
\newcommand{\xigg}{$\xi_{\rm gg}$\xspace}
\newcommand{\kms}{\,km s$^{-1}$\xspace}
\newcommand{\cm}{\,cm$^{-2}$\xspace}
\newcommand{\mpc}{\,Mpc\xspace}

\title[The intergalactic medium in the cosmic web] 
{The intergalactic medium in the cosmic web}

\author[Nicolas Tejos] 
{Nicolas Tejos\footnote{On behalf of our full collaboration.}
}

\affiliation{Department of Astronomy and Astrophysics,\\
UCO/Lick Observatory, University of California,\\ 
1156 High Street, Santa Cruz, CA 95064, USA\\
email: {\tt ntejos@ucolick.org}}

\pubyear{2014}
\volume{308}  
\pagerange{119--126}
 \jname{The Zeldovich Universe: Genesis and Growth
  of the Cosmic Web} \editors{Rien van de Weygaert, Sergei Shandarin,\\
  Enn Saar and Jaan Einasto}
\begin{document}

\maketitle

\begin{abstract}
The intergalactic medium (IGM) accounts for $\gtrsim90\%$ of baryons at
all epochs and yet its three dimensional distribution in the cosmic web
remains mostly unknown. This is so because the only feasible way to
observe the bulk of the IGM is through intervening absorption line
systems in the spectra of bright background sources, which limits its
characterization to being one-dimensional. Still, an averaged three
dimensional picture can be obtained by combining and cross-matching
multiple one-dimensional IGM information with three-dimensional galaxy
surveys. Here, we present our recent and current efforts to map and
characterize the IGM in the cosmic web using galaxies as tracers of the
underlying mass distribution. In particular, we summarize our results
on: (i) IGM around star-forming and non-star-forming galaxies; (ii) IGM
within and around galaxy voids; and (iii) IGM in intercluster
filaments. With these datasets, we can directly test the modern
paradigm of structure formation and evolution of baryonic matter in the
Universe.

\end{abstract}

\firstsection 
\section{Introduction}

The physics of the intergalactic medium (IGM) and its connection with
galaxies are key to understanding the evolution of baryonic matter in
the Universe. The IGM is the main reservoir of baryons at all epochs
(e.g. \cite[Fukugita et al. 1998]{Fukugita1998}; \cite[Shull et
  al. 2012]{Shull2012}), and provides the primordial material for
forming galaxies. Once galaxies are formed, supernovae (SNe) and
active-galactic nuclei (AGN) feedback inject energy in the interstellar
medium, some of which escapes the galaxies as winds, enriching the IGM
with metals (e.g. \cite[Wiersma et al. 2011]{Wiersma2011}; \cite[Ford
  et al. 2014]{Ford2014}). Because of the continuous interplay between
the IGM and galaxies, it is sensible (if not necessary) to study these
two concepts simultaneously (e.g. \cite[Morris et
  al. 1993]{Morris1993}; \cite[Lanzetta et al. 1995]{lanzetta1995};
\cite[Tripp et al. 1998]{Tripp1998}; \cite[Chen \& Mulchaey
  2009]{Chen2009}; \cite[Prochaska et al. 2011]{Prochaska2011};
\cite[Tumlinson et al. 2011]{Tumlinson2011}; \cite[Tejos et
  al. 2014]{Tejos2014}; \cite[Werk et al. 2014]{Werk2014}).\smallskip

The large scale environment in which matter resides also plays an
important role. Given that baryonic matter is expected to fall into the
considerably deeper gravitational potentials of dark matter, the IGM
gas and galaxies should be predominantly found at such locations,
forming the so-called \lq cosmic web\rq\ (\cite[Bond et
  al. 1996]{Bond1996}). Galaxies appear to follow the filamentary
structure which simulations predict (e.g. \cite[Springel et
  al. 2006]{Springel2006}), and their properties are partly shaped by
environmental effects (e.g. \cite[Dressler 1980]{Dressler1980};
\cite[Skibba et al. 2009]{Skibba2009}). However, much less is known
about the {\it actual} properties and distribution of the IGM in
different cosmological environments. This is so because the only
feasible way to observe the bulk of the IGM is through intervening
absorption line systems in the spectra of bright background sources
(e.g. quasi-stellar objects, gamma-ray bursts, galaxies), which limits
its characterization to being one-dimensional.

The advent of big galaxy surveys such as the 2dFGRS (\cite[Colless et
  al. 2001]{Colless2001}) or the SDSS (\cite[Abazajian et
  al. 2009]{Abazajian2009}), have revolutionized the study of the
cosmic web and the large-scale structure (LSS) of the Universe. This is
eloquently demonstrated by the plethora of LSS catalogs that are
currently available: from galaxy voids (e.g. \cite[Pan et
  al. 2012]{Pan2012}; \cite[Sutter et al. 2012]{Sutter2012};
\cite[Nadathur \& Hotchkiss 2014]{Nadathur2014}; \cite[Way et
  al. 2014]{Way2014}), galaxy filaments (e.g. \cite[Tempel et
  al. 2014]{Tempel2014}), to galaxy groups and clusters (e.g. \cite[Hao
  et al. 2010]{Hao2012}; \cite[Rykoff et al. 2014]{Rykoff2014}). By
combining and cross-matching multiple one-dimensional IGM information
with galaxy and LSS surveys, an averaged three dimensional picture can
be obtained. Here, we present our recent and current efforts to map and
characterize the IGM in the cosmic web using galaxies as tracers of the
underlying mass distribution.

\begin{figure}[t]
 \vspace*{-0.1 cm}
\begin{center}
 \includegraphics[width=4.0in]{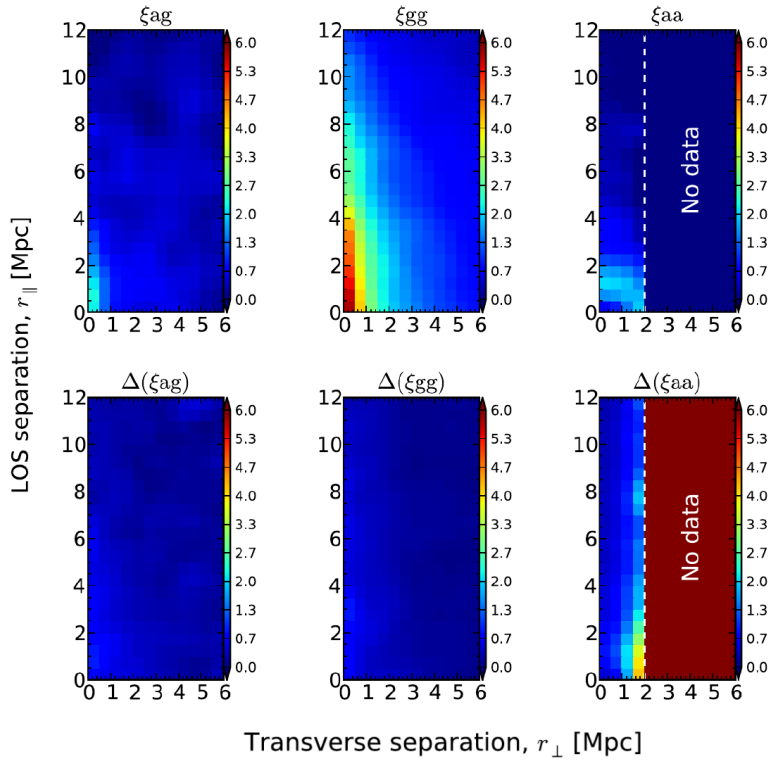}
 \caption{Two-dimensional two-point correlation functions (top panels)
   at $z\lesssim 1$, and their respective uncertainties (bottom
   panels). From left to right: \hi--galaxy (\xiag), galaxy--galaxy
   (\xigg) and \hi--\hi (\xiaa) cross-correlations. Figure adapted from
   \cite[Tejos et al. (2014)]{Tejos2014}.}
   \label{fig:1}
\end{center}
\end{figure}

\section{The IGM-galaxy cross-correlation}

The two-point correlation function between neutral hydrogen (\hi) and
galaxies is a powerful statistical technique to assess the connection
between the IGM and galaxies (e.g. \cite[Chen et al. 2005]{Chen2005};
\cite[Ryan-Weber 2006]{Ryan-Weber2006}; \cite[Wilman et
  al. 2007]{Wilman2007}; \cite[Chen \& Mulchaey 2009]{Chen2009};
\cite[Shone et al. 2010]{Shone2010}; \cite[Tejos et
  al. 2014]{Tejos2014}).

In \cite[Tejos et al. (2014)]{Tejos2014}, we have recently published
observational results on the \hi--galaxy two-point cross-correlation at
$z\lesssim 1$ (\xiag; see Fig.\,\ref{fig:1}). These results come from
the largest sample ever done for such an analysis, comprising about
$\sim 700$ \hi absorption line systems in the UV spectra of $8$
background QSOs, in $6$ different fields observed with the HST, and
about $\sim 17000$ galaxies with spectroscopic redshifts around these
QSOs sightlines, coming from our own spectroscopic surveys, and
previously published catalogs by the VVDS (\cite[Le F{\`e}vre et
  al. 2013]{LeFevre2013}) and GDDS (\cite[Abraham et
  al. 2004]{Abraham2004}) galaxy surveys.

Apart from \xiag, we also measured the \hi--\hi (\xiaa) and
galaxy-galaxy (\xigg) two-point auto-correlations. Our survey is one of
the few in which these three quantities have been measured from the
same dataset, and independently from each other. Comparing the results
from \xiag, \xiaa and \xigg, we constrained the IGM-galaxy statistical
connection, as a function of both \hi~column density and galaxy star
formation activity, on $\sim 0.5 - 10$\mpc scales. Our results are
consistent with the following conclusions: (i) the bulk of \hi~systems
on $\sim$\mpc scales have little velocity dispersion ($\lesssim
120$\kms) with respect to the bulk of galaxies (i.e. no strong galaxy
outflow/inflow signal is detected); (ii) the vast majority
($\sim100\%$) of \hi systems with $N_{\rm HI}>10^{14}$\cm and
star-forming galaxies are distributed in the same locations, together
with $75 \pm 15\%$ of non-star forming galaxies; (iii) $25 \pm 15\%$ of
non-star-forming galaxies reside in galaxy clusters and are not
correlated with \hi~systems at scales $\lesssim 2$\mpc; and (iv)
$>50\%$ of \hi systems with $N_{\rm HI}<10^{14}$\cm reside within
galaxy voids and hence are not correlated with luminous galaxies.

\section{The IGM within and around galaxy voids}

In \cite[Tejos et al. (2012)]{Tejos2012} we have recently measured the
properties of \hi absorption line systems within and around galaxy
voids at $z\le 0.1$, using the galaxy void catalog published by
\cite[Pan et al. (2011)]{Pan2011} and the low-$z$ \hi~absorption line
catalog published by \cite[Danforth \& Shull (2008)]{Danforth2008}. Our
key findings can be summarized as follows: (i) there is a significant
excess of IGM gas at the edges of galaxy voids with respect to the
random expectation, {\it consistent} with the overdensity of galaxies
defining such voids; and (ii) inside galaxy voids the IGM gas matches
the random expectation, {\it inconsistent} with the underdensity of
galaxies defining such voids. In other words, there were no apparent
IGM voids detected at the positions of galaxy voids.

We also showed that the column density ($N_{\rm HI}$) and Doppler
parameter ($b_{\rm HI}$) distributions of \hi\ lines inside and outside
galaxy voids were not remarkably different, with only a $\sim 95\%$ and
$\sim 90\%$ probability of rejecting the null-hypothesis of both
samples coming from the same parent population, respectively. Still, a
trend was present, in which galaxy void absorbers have systematically
lower values of both $N_{\rm HI}$ and $b_{\rm HI}$ than those found
outside galaxy voids. By performing a similar analysis using a
state-of-the-art hydrodynamical cosmological simulation (GIMIC;
\cite[Crain et al. 2009]{Crain2009}), we showed that these observed
trends are qualitatively consistent with current theoretical
expectations. However, more quantitative comparisons of the gas
properties in galaxy voids between hydrodynamical simulations tuned to
explore low density environments (e.g. \cite[Ricciardelli et
  al. 2013]{Ricciardelli2013}) and observations (e.g. \cite[Tejos et
  al. 2012]{Tejos2012}), are required.

\section{The IGM in intercluster filaments}
Galaxy clusters represent the densest nodes in the cosmic web (with
dark matter halo masses of $M\gtrsim10^{14}M_{\odot}$) and as such,
$N$-body numerical simulations predict a high probability of finding
intercluster filaments between galaxy cluster pairs when separated by
$\lesssim10 - 20$\mpc (e.g. \cite[Colberg et al. 2005]{Colberg2005};
\cite[Gonz{\'a}lez \& Padilla 2010]{Gonzalez2010}). Hydrodynamical
simulations predict that an important fraction of baryons at low-$z$
are in a diffuse, shock heated gas phase with $T\sim 10^{5}-10^{6}$\,K
in these dense filaments, commonly referred to as the warm-hot
intergalactic medium (WHIM; \cite[Cen \& Ostriker 1999]{Cen1999};
\cite[Dav{\'e} et al. 2001]{Dave2001}). However, this WHIM has been
very elusive and difficult to observe (e.g. \cite[Richter et
  al. 2006]{Richter2006}).

Here, we have presented preliminary results on the properties of the
IGM in intercluster filaments at $0.1 \le z \le 0.47$ by using a single
QSO observed with the HST/COS UV spectrograph, whose sightline
intersects $7$ independent cluster-pairs at impact parameters $<5$\mpc
from the intercluster axes. This technique allowed us to perform, for
the first time, a systematic and statistical measurement of the
incidence of \hi\ and \ovi\ absorption lines associated to intercluster
filaments. We constrained the geometry and physical properties of the
IGM gas lying between clusters. Our results are consistent with a
filamentary geometry for the gas, and the presence of both broad \hi
($> 50$\kms) and \ovi hint towards the existence of a WHIM (Tejos et
al. in prep.).\smallskip\smallskip

{\it This work was partly funded by CONICYT/PFCHA 72090883 (Chile).}



\vspace*{-0.4 cm}
\begin{thebibliography}{}  

\bibitem[Abraham et al.(2004)]{Abraham2004} Abraham, R.~G., Glazebrook,
  K., McCarthy, P.~J., et al.\ 2004, \aj, 127, 2455

\bibitem[Abazajian et al.(2009)]{Abazajian2009} Abazajian, K.~N.,
  Adelman-McCarthy, J.~K., Ag{\"u}eros, M.~A., et al.\ 2009, \apjs,
  182, 543

\bibitem[Bond et al.(1996)]{Bond1996} Bond, J.~R., Kofman, L., \&
  Pogosyan, D.\ 1996, \nat, 380, 603

\bibitem[Cen \& Ostriker(1999)]{Cen1999} Cen, R., \& Ostriker,
  J.~P.\ 1999, \apj, 514, 1

\bibitem[Chen et al.(1998)]{Chen1998} Chen, H.-W., Lanzetta, K.~M.,
  Webb, J.~K., \& Barcons, X.\ 1998, \apj, 498, 77

\bibitem[Chen et al.(2005)]{Chen2005} Chen, H.-W., Prochaska, J.~X.,
  Weiner, B.~J., et al.\ 2005, \apj (Letters), 629, L25

\bibitem[Chen \& Mulchaey(2009)]{Chen2009} Chen, H.-W., \& Mulchaey,
  J.~S.\ 2009, \apj, 701, 1219

\bibitem[Colberg et al.(2005)]{Colberg2005} Colberg, J.~M., Krughoff,
  K.~S., \& Connolly, A.~J.\ 2005, \mnras, 359, 272

\bibitem[Colless et al.(2001)]{Colless2001} Colless, M., Dalton, G.,
  Maddox, S., et al.\ 2001, \mnras, 328, 1039

\bibitem[Crain et al.(2009)]{Crain2009} Crain, R.~A., Theuns, T., Dalla
  Vecchia, C., et al.\ 2009, \mnras, 399, 1773

\bibitem[Danforth \& Shull(2008)]{Danforth2008} Danforth, C.~W., \&
  Shull, J.~M.\ 2008, \apj, 679, 194

\bibitem[Dav{\'e} et al.(2001)]{Dave2001} Dav{\'e}, R., Cen, R.,
  Ostriker, J.~P., et al.\ 2001, \apj, 552, 473

\bibitem[Dressler(1980)]{Dressler1980} Dressler, A.\ 1980, \apj, 236,
  351

\bibitem[Ford et al.(2014)]{Ford2014} Ford, A.~B., Dav{\'e},
  R., Oppenheimer, B.~D., et al.\ 2014, \textit{MNRAS}, 444, 1260

\bibitem[Fukugita et al.(1998)]{Fukugita1998} Fukugita, M., Hogan,
  C.~J., \& Peebles, P.~J.~E.\ 1998, \textit{ApJ}, 503, 518

\bibitem[Gonz{\'a}lez \& Padilla(2010)]{Gonzalez2010} Gonz{\'a}lez,
  R.~E., \& Padilla, N.~D.\ 2010, \mnras, 407, 1449

\bibitem[Hao et al.(2010)]{Hao2010} Hao, J., McKay, T.~A., Koester,
  B.~P., et al.\ 2010, \apjs, 191, 254

\bibitem[Lanzetta et al.(1995)]{Lanzetta1995} Lanzetta, K.~M., Bowen,
  D.~V., Tytler, D., \& Webb, J.~K.\ 1995, \apj, 442, 538


\bibitem[Le F{\`e}vre et al.(2013)]{LeFevre2013} Le F{\`e}vre, O.,
  Cassata, P., Cucciati, O., et al.\ 2013, \aap, 559, A14

\bibitem[Morris et al.(1993)]{Morris1993} Morris, S.~L., Weymann,
  R.~J., Dressler, A., et al.\ 1993, \apj, 419, 524

\bibitem[Nadathur \& Hotchkiss(2014)]{Nadathur2014} Nadathur, S., \&
  Hotchkiss, S.\ 2014, \mnras, 440, 1248

\bibitem[Pan et al.(2012)]{Pan2012} Pan, D.~C., Vogeley, M.~S., Hoyle,
  F., Choi, Y.-Y., \& Park, C.\ 2012, \mnras, 421, 926

\bibitem[Prochaska et al.(2011)]{Prochaska2011} Prochaska, J.~X.,
  Weiner, B., Chen, H.-W., Mulchaey, J., \& Cooksey, K.\ 2011, \apj,
  740, 91

\bibitem[Ricciardelli et al.(2013)]{Ricciardelli2013} Ricciardelli, E.,
  Quilis, V., \& Planelles, S.\ 2013, \mnras, 434, 1192

\bibitem[Richter et al.(2006)]{Richter2006} Richter, P., Savage, B.~D.,
  Sembach, K.~R., \& Tripp, T.~M.\ 2006, \aap, 445, 827

\bibitem[Ryan-Weber(2006)]{Ryan-Weber2006} Ryan-Weber, E.~V.\ 2006,
  \mnras, 367, 1251

\bibitem[Rykoff et al.(2014)]{Rykoff2014} Rykoff, E.~S., Rozo, E.,
  Busha, M.~T., et al.\ 2014, \apj, 785, 104

\bibitem[Shone et al.(2010)]{Shone2010} Shone, A.~M., Morris, S.~L.,
  Crighton, N., \& Wilman, R.~J.\ 2010, \mnras, 402, 2520

\bibitem[Shull et al.(2012)]{Shull2012} Shull, J.~M., Smith, B.~D., \&
  Danforth, C.~W.\ 2012, \textit{ApJ}, 759, 23

\bibitem[Skibba et al.(2009)]{Skibba2009} Skibba, R.~A., Bamford,
  S.~P., Nichol, R.~C., et al.\ 2009, \mnras, 399, 966

\bibitem[Springel et al.(2006)]{Springel2006} Springel, V., Frenk,
  C.~S., \& White, S.~D.~M.\ 2006, \nat, 440, 1137

\bibitem[Sutter et al.(2012)]{Sutter2012} Sutter, P.~M., Lavaux, G.,
  Wandelt, B.~D., \& Weinberg, D.~H.\ 2012, \apj, 761, 44

\bibitem[Tejos et al.(2012)]{Tejos2012} Tejos, N., Morris, S.~L.,
  Crighton, N.~H.~M., et al.\ 2012, \mnras, 425, 245

\bibitem[Tejos et al.(2014)]{Tejos2014} Tejos, N., Morris, S.~L., Finn,
  C.~W., et al.\ 2014, \mnras, 437, 2017

\bibitem[Tempel et al.(2014)]{Tempel2014} Tempel, E., Stoica, R.~S.,
  Mart{\'{\i}}nez, V.~J., et al.\ 2014, \mnras, 438, 3465

\bibitem[Tripp et al.(1998)]{Tripp1998} Tripp, T.~M., Lu, L., \&
  Savage, B.~D.\ 1998, \apj, 508, 200

\bibitem[Tumlinson et al.(2011)]{Tumlinson2011} Tumlinson, J., Thom,
  C., Werk, J.~K., et al.\ 2011, \textit{Science}, 334, 948

\bibitem[Way et al.(2014)]{Way2014} Way, M.~J., Gazis, P.~R., \&
  Scargle, J.~D.\ 2014, arXiv:1406.6111

\bibitem[Werk et al.(2014)]{Werk2014} Werk, J.~K., Prochaska, J.~X.,
  Tumlinson, J., et al.\ 2014, \apj, 792, 8

\bibitem[Wiersma et al.(2011)]{Wiersma2011} Wiersma, R.~P.~C., Schaye,
  J., \& Theuns, T.\ 2011, \textit{MNRAS}, 415, 353

\bibitem[Wilman et al.(2007)]{Wilman2007} Wilman, R.~J., Morris, S.~L.,
  Jannuzi, B.~T., Dav{\'e}, R., \& Shone, A.~M.\ 2007, \mnras, 375, 735





\end{thebibliography}
\end{document}